\documentclass[sn-mathphys-num]{sn-jnl}

\usepackage{graphicx}%
\usepackage{multirow}%
\usepackage{amsmath,amssymb,amsfonts}%
\usepackage{amsthm}%
\usepackage{mathrsfs}%
\usepackage[title]{appendix}%
\usepackage{xcolor}%
\usepackage{textcomp}%
\usepackage{manyfoot}%
\usepackage{booktabs}%
\usepackage{algorithm}%
\usepackage{algorithmicx}%
\usepackage{algpseudocode}%
\usepackage{listings}%
\usepackage{orcidlink}%

\theoremstyle{thmstyleone}%
\theoremstyle{thmstyletwo}%

\theoremstyle{thmstylethree}%

\begin{document}

\title[Shear Viscosity of Collider-Produced QCD Matter II: Comparing a Multi-Component Chapman-Enskog Framework with AMPT in Full Equilibrium]{Shear Viscosity of Collider-Produced QCD Matter II: Comparing a Multi-Component Chapman-Enskog Framework with AMPT in Full Equilibrium}


\author[1]{\fnm{Noah M.} \sur{MacKay}\,\orcidlink{0000-0001-6625-2321}}\email{noah.mackay@uni-potsdam.de}

\affil[1]{\orgdiv{Institute of Physics and Astronomy}, \orgname{Universit\"at Potsdam}, \orgaddress{\street{Karl-Liebknecht-Stra\ss e 24/25}, \postcode{14476} \city{Potsdam}, \country{Germany}}}


\abstract{Transport properties of the quark-gluon plasma are instrumental to testing perturbative quantum chromodynamics and understanding the extreme conditions of relativistic heavy-ion collisions. This study presents an analytical investigation of the shear viscosity $\eta$ and the shear viscosity-to-entropy density ratio $\eta/s$ of the QGP using a novel multi-component Chapman-Enskog framework assuming full thermalization. The approach incorpoartes species-specific contributions from gluons and (anti-)quarks into the plasma shear viscosity, temperature-dependent running parameters for the Debye mass and strong coupling, and a time-dependent cooling model. Our findings show that both $\eta$ and $\eta/s$ are enhanced by the inclusion of (anti-)quarks with gluons, and the parameters decrease over time due to the cooling and expansion of the QGP. These results align with perturbative QCD predictions, offering a more optimistic representation of QGP transport properties under dynamic conditions. This multi-component framework is compared with a multi-phase transport model that treats the QGP as a gluon gas with (anti-)quark augmentation.}




\maketitle

\section{Introduction}\label{sec1}

Relativistic heavy-ion collisions, such as those at the Relativistic Heavy Ion Collider (RHIC) and the Large Hadron Collider (LHC) \cite{J.Adams:STAR, PHENIX:2004vcz}, create a quark-gluon plasma (QGP). The QGP is an extremely hot and dense system where (anti-)quarks and gluons exist as deconfined, quasi-free partons. Past comparisons between experimental anisotropic flow measurements and theoretical models (i.e., hydrodynamics \cite{Romatschke:2007mq, Song:2008si, ALICE:2010suc} and kinetic transport methods \cite{Lin:2001zk, Xu:2007jv, Xu:2008av, Ferini:2008he}) suggest that the QGP behaves like a near-perfect fluid with a very small shear-viscosity to entropy-density ratio above the KSS lower bound $\eta/s= 1/(4\pi)$ ($\hbar=k_B=c=1$) \cite{Kovtun:2004de}. At temperatures exceeding the Hagedorn temperature, $T_H\simeq150$ MeV \cite{Gazdzicki:2015oya}, the partons in the QGP resemble an ideal thermal gas. Natural units are used throughout this study.

Both numerical hydrodynamic and transport models describe the shear viscosity $\eta$ of the QGP, albeit using different approaches. In hydrodynamic models, the $\eta/s$ ratio (and its temperature dependence) is treated as an input parameter. In transport models, viscosity is derived from temperature-dependent scattering cross sections \cite{Xu:2007jv, Ferini:2008he, Xu:2007ns, Xu:2011fi}. Analytically, approaches in kinetic theory relate viscosity with the account of particle interactions in the linearized collision kernel $\mathcal{C}[f]$ (from the integro-differential Boltzmann equation $p^\mu\partial_\mu f(p)=\mathcal{C}[f]$) via particle distribution functions $f(p)$, their relative motions, and their shared differential cross section.

Two prominent methods for determining $\eta$ are the numerical Green-Kubo method, which extracts $\eta$ from equilibrium correlation functions \cite{Kovtun:2004de, Plumari:2012ep, MacKay:2022uxo}, and the analytical Chapman-Enskog (CE) method, which solves the Boltzmann equation by applying the method of moments to the distribution function \cite{ce, degroot}. In these studies, the partons in the QGP are often treated as massless, ultra-relativistic particles obeying Maxwell-Boltzmann (MB) statistics. This simplification is valid in the high-temperature regime, where quantum statistical effects become negligible. For collider-produced QGP, the CE method consistently agrees with numerical results for both isotropic and anisotropic scattering cases \cite{Plumari:2012ep, MacKay:2022uxo}. Furthermore, temperature plays a crucial role in scattering anisotropy, as demonstrated in Ref. \cite{mackay1}. At higher temperatures, scattering anisotropy increases, leading to an enhanced viscosity. Conversely, isotropic scatterings dominate in lower temperatures, though these conditions fall outside the deconfined parton phase of the QGP. 

The agreement between the CE and Green-Kubo methods facilitated advanced studies of QGP dynamics, such as those in Ref. \cite{MacKay:2022uxo} that examined the time evolution of the QGP using a multi-phase transport (AMPT) model. The AMPT model is a hybrid model incorporating partonic- and hadronic-transport processes to generate heavy-ion collision events, with parton interactions based on the two-body forward-angle scattering process of the Zhang parton cascade (ZPC) model \cite{Zhang:1997ej}. The ZPC model represents the QGP as an effective gluon gas augmented by $N_f$-flavor quark degrees of freedom \cite{Plumari:2012ep, MacKay:2022uxo, Zhang:1997ej, Lin:2014tya, Wang:2021owa}. The CE expressions for anisotropic $\eta$ and $\eta/s$ were semi-analytically calculated for two cases: full equilibrium and partial equilibrium (i.e., the system is thermalized but not chemically equilibrated). These calculations were tracked over the period when the QGP remains in a deconfined parton state, approximately 10 fm after heavy-ion collision \cite{Lin:2014tya}. To model parton interactions, the AMPT and ZPC cross sections is based on the pQCD gluon-gluon scattering amplitude (c.f. Ref. \cite{Arnold:2003zc}, Appendix A). 

A recent study, which compared a multi-component hydrodynamic expression with single-component kinetic transport model calculations for shear viscsoity under isotropic scatterings, demonstrated that a standard one-component description is generally insufficient for multi-component systems \cite{El:2011cp}. Multi-component (i.e., $N$-component) shear viscosity with nonzero particle mass has been studied for polyatomic mixtures using the Sutherland formula \cite{nasa, suth}, and studies on nonpolar mixtures utilized the CE method \cite{degroot, Moroz:2014}. However, these $N$-component CE viscosity studies only considered the simpler yet specific case of isotropic scatterings. Multi-component shear viscosity under general-case anisotropic scatterings were focused on massless, Maxwell-distributed particles in Ref. \cite{MacKay:2024apd}. Extending these approaches to a multi-component QGP requires analysis beyond the gluon gas simplification to incorporate (anti-)quarks.

The ultra-relativistic behavior of partons in the QGP leads to a small strong coupling constant due to asymptotic freedom, as the interaction strength diminishes at high energy scales \cite{Gross:1973id}. This allows the scattering amplitude $|\mathcal{M}|^2$ to directly determine the differential cross section via $d\sigma/d\Omega\propto|\mathcal{M}|^2$, with phase-space factors depending on the initial and final states. Neglecting their masses, (anti-)quarks become kinematically flavor-independent, allowing the interactions to be categorized as same-fermion and different-fermion interactions. As a result, the relevant flavors are captured in the fermion degrees of freedom, and the QGP can be effectively modeled as a three-component system of gluons ($g$), quarks ($q$), and antiquarks ($\bar{q}$). 

\section{Methods}

\subsection{$N$-Component Chapman-Enskog Viscosity}

The shear viscosity for an $N$-component system, using the Chapman-Enskog method, is expressed as an iterative sum of partial viscosities between species $l$ and $k$ \cite{degroot, Moroz:2014}:
\begin{equation}
\begin{split}
&\eta=\frac{T}{10\sigma}\sum_{k=1}^{N}x_k\gamma_{0,k}C_{0,k},~~~~~\gamma_{0,k}x_k=\sum_{l=1}^N C_{0,l}C^{00}_{lk},\\
&\gamma_0=-10\frac{K_3(z)}{K_2(z)},~~~~~x_k=\frac {n_k}{\sum_{k'}^N n_{k'}}.
\end{split}
\end{equation}
Here, $K_n$ is the modified Bessel function of the second kind, with $z=m/T$ as the dimensionless mass parameter. The molar fraction $x_k$ satisfies $\sum_{k=1}^N x_k=1$, where $n_k$ is the quantum number density of the $k$-th particle species and $\sum_{k'=1}^N n_{k'}$ represents the total number density. Furthermore, $\sigma=\sum_{l,k}^N\sigma_{lk}$ represent the grand total cross section, which is the sum of the total cross sections for all interactions. For the case of massless MB-distributed partons with spin-color degrees of freedom, their respective quantum number densities are determined by MB statistics and weighted by their respective degeneracy factors. Also, $\gamma_0=1600/z^2$ for asymptotically small $z\propto m$.

The factor $C_{0,k}$ in the Chapman-Enskog framework represents the contribution of the $k$-th species in the total shear viscosity. It depends on the molar fraction $x_k$, the enthalpy factor $\gamma_{0,k}$, and the linearized collision kernel $C^{00}_{lk}$ that quantifies the strength of the elastic $2\leftrightarrow2$ scatterings between species $l$ and $k$. For $N=3$, iterative expansions on $C_{0,j}$ have been derived in Ref. \cite{MacKay:2024apd}, under the assumption that $\gamma_{0,j}=\gamma_0$ is identical for all massless species, and are given by:
\begin{equation}
\begin{split}
& C_{0,1}=\frac{\gamma_{0}x_1}{C^{00}_{11}},\\
& C_{0,2}=\left[x_{2}-x_1\frac{2C^{00}_{12}}{C^{00}_{11}} \right]\frac{\gamma_0}{C^{00}_{22}},\\
&C_{0,3}= \gamma_0\left[\frac{x_3}{C^{00}_{33}}-\frac{2x_1C^{00}_{13}}{C^{00}_{33}C^{00}_{11}}-\left(\frac{x_2}{C^{00}_{22}}-\frac{2x_1C^{00}_{12}}{C^{00}_{22}C^{00}_{11}} \right) \left(\frac{2C^{00}_{23}}{C^{00}_{33}}+\frac{C^{00}_{12}}{C^{00}_{13}} \right)  \right].
\end{split}
\end{equation}
In the above definitions, the linearized collision kernel has the symmetric property $C^{00}_{lk}=C^{00}_{kl}$. The total shear viscosity for an $N=3$ system is therefore expressed as
\begin{equation} \label{visc}
\begin{split}
\eta=\frac{T\gamma_0}{10\sigma}\left(x_1C_{0,1}+x_2C_{0,2}+x_3C_{0,3}\right).
\end{split}
\end{equation}

The linearized collision kernel $C^{00}_{lk}$ for massless, MB-distributed species $l$ and $k$ is derived under anisotropic scatterings in Ref. \cite{MacKay:2024apd} and expressed as
\begin{equation} \label{anisodg}
\begin{split}
C^{00}_{kl}&\simeq\frac{5^{\delta_{N,1}}}{32z^2 \sigma}\frac{\left[\pi^{(\delta_{lk}-1)}2^{(2-\delta_{lk})}\right]^{\Delta_{N,1}}}{(11-6\delta_{kl})}\left[x_kx_l-2\delta_{kl}x_k\left(\frac{7}{4}\right)^{\delta_{kl}\Delta_{N,1}}\sum_{m=1}^Nx_m \right]\\
&~~~\times\Big\{\frac{(\delta_{kl}-1)}{2}T_{lk}+(\delta_{kl}-2)J_{lk} \Big\},
\end{split}
\end{equation}
where $\delta_{ij}$ is the Kronecker delta; $\delta_{N,1}$ applies to orthonormality between the total number of species and the case of $N=1$, and $\Delta_{N,1}=1-\delta_{N,1}$. The terms $T_{lk}$ and $J_{lk}$ are integral expressions that depend on the Mandelstam variables $\hat{s}$ and $\hat{t}$, as well as kinematic weightings on the differential cross section $d\sigma_{lk}/d\hat{t}$:
\begin{equation}
\begin{split}
&T_{lk}=\frac{1}{2T^9}\int_0^\infty d\hat{s}~\hat{s}^{7/2}K_1\left(\frac{\sqrt{\hat{s}}}{T}\right)\int_{-\hat{s}}^0 d\hat{t}\left(\frac{8}{3}+\frac{4\hat{t}}{\hat{s}}\right)\frac{d\sigma_{lk}}{d\hat{t}},\\\\
&J_{lk}=\frac{1}{2T^7}\int_0^\infty d\hat{s}~\hat{s}^{5/2} \left[\left(\frac{\hat{s}}{4T^2}+\frac{1}{3}\right)K_3\left(\frac{\sqrt{\hat{s}}}{T}\right)-\frac{\sqrt{\hat{s}}}{2T}K_2\left(\frac{\sqrt{\hat{s}}}{T}\right) \right]\\
&~~~~~~~~~~~~~\times\int_{-\hat{s}}^0 d\hat{t}\left(\frac{-4\hat{t}^2}{\hat{s}^2}-\frac{4\hat{t}}{\hat{s}}\right)\frac{d\sigma_{lk}}{d\hat{t}}.
\end{split}
\end{equation}
 For like-species interactions ($l=k$), the first integral $T_{lk}$ vanishes, simplifying the overall expression. The differential cross section is differentiated with respect to the $\hat{t}$-Mandelstam variable, which is related to the solid-angle differential cross section as 
\begin{equation}
\frac{d\sigma_{lk}}{d\hat{t}}=\frac{4\pi}{\hat{s}}\frac{d\sigma_{lk}}{d\Omega}.
\end{equation}
Here, $\hat{s}$ is the center-of-momentum energy of the incoming particles, and $d\sigma_{lk}/d\Omega$ is given in terms of the Mandelstam variables $\hat{s}$, $\hat{t}$, and $\hat{u}$. In the viscosity expression (Eq. [\ref{visc}]), the grand total cross section $\sigma=\sum_{l,k}\sigma_{lk}$ cancels out, leaving the viscosity dependent on the specific interaction cross sections $\sigma_{lk}$ and their kinematic weightings.

\subsection{Elastic $2\leftrightarrow2$ Parton Interactions}

In an three-component QGP, gluons and (anti-)quarks interact through self-interactions ($gg\leftrightarrow gg$, $qq\leftrightarrow qq$, and $\bar{q}\bar{q}\leftrightarrow\bar{q}\bar{q}$) and cross-species interactions ($gq\leftrightarrow gq$, $g\bar{q}\leftrightarrow g\bar{q}$, and $q\bar{q}\leftrightarrow q\bar{q}$). Ref. \cite{rpp2021} provides a comprehensive list of Standard Model cross section formulas that include the relevant $N=3$ QGP interactions:
\begin{equation}
\frac{d\sigma}{d\Omega}(q\bar{q}\leftrightarrow q\bar{q})=\frac{\alpha_s^2}{9\hat{s}}\left[\frac{\hat{t}^2+\hat{u}^2}{\hat{s}^2}+\frac{\hat{s}^2+\hat{u}^2}{\hat{t}^2}-\frac{2\hat{u}^2}{3\hat{s}\hat{t}} \right],
\end{equation}
\begin{equation}
\frac{d\sigma}{d\Omega}(q{q}\leftrightarrow q{q}~~\mathrm{and}~~\bar{q}\bar{q}\leftrightarrow \bar{q}\bar{q})=\frac{\alpha_s^2}{9\hat{s}}\left[\frac{\hat{t}^2+\hat{u}^2}{\hat{s}^2}+\frac{\hat{s}^2+\hat{u}^2}{\hat{t}^2}-\frac{2\hat{s}^2}{3\hat{u}\hat{t}} \right],
\end{equation}
\begin{equation}
\frac{d\sigma}{d\Omega}(g{q}\leftrightarrow g{q}~~\mathrm{and}~~g\bar{q}\leftrightarrow g\bar{q})=\frac{\alpha_s^2}{9\hat{s}}\left(\hat{s}^2+\hat{u}^2\right)\left[-\frac{1}{\hat{s}\hat{u}}+\frac{9}{4\hat{t}^2} \right],
\end{equation}
\begin{equation}
\frac{d\sigma}{d\Omega}(gg\leftrightarrow gg)=\frac{9\alpha_s^2}{8\hat{s}}\left[3-\frac{\hat{u}\hat{t}}{\hat{s}^2}-\frac{\hat{s}\hat{u}}{\hat{t}^2}-\frac{\hat{s}\hat{t}}{\hat{u}^2} \right].
\end{equation}
Here, $\alpha_s$ is the strong coupling constant, and $\hat{u}=-\hat{t}-\hat{s}$ for the case of massless particles. To avoid overcounting of identical interactions into the total shear viscosity, we assign index labels to the parton species: $g=1,~q=2,~\bar{q}=3$. The elastic $l+k\leftrightarrow l+k$ interactions in the $N=3$ QGP are summarized in Table \ref{tab5}, grouped by identical/different scattering amplitudes.

\begin{table}[h]
\caption{Elastic $l+k\rightarrow l+k$ interactions in the $N=3$ QGP}\label{tab5}
\begin{tabular}{l|llll}
\toprule
Parton Interactions & $gg$ & $gq$ & $qq$ &  $q\bar{q}$  \\
& & $g\bar{q}$ & $\bar{q}\bar{q}$ &    \\
\midrule
$l,k$ Index Labels for $C^{00}_{lk}$ &11 & 12,~21 & 22 & 23,~32  \\
 
&  & 13,~31 & 33 &    \\
 \midrule
Simplified Index Labels&11 & 12 & 22 & 23  \\
\botrule
\end{tabular}
\end{table}

Given that the (anti-)quarks share a common number density ($x_2=x_3$), the iterative expansions for $C_{0,j}$ simplify significantly:
\begin{equation}
\begin{split}
& C_{0,1}=\frac{\gamma_{0}x_1}{C^{00}_{11}},~~~ C_{0,2}=\frac{\gamma_0}{C^{00}_{22}}\left[x_{2}-x_1\frac{2C^{00}_{12}}{C^{00}_{11}} \right],~~~C_{0,3}= C_{0,2}\left[-\frac{2C^{00}_{23}}{C^{00}_{22}}  \right].
\end{split}
\end{equation}
Therefore, the shear viscosity for the $N=3$ QGP, incorporating these simplifications, is given by
\begin{equation} \label{viscsimp}
\begin{split}
\eta=\frac{T\gamma^2_0}{10\sigma}\left[\frac{x^2_1}{C^{00}_{11}}+\left(\frac{x^2_{2}}{C^{00}_{22}}-x_1x_2\frac{2C^{00}_{12}}{C^{00}_{22}C^{00}_{11}} \right)\left(1-\frac{2C^{00}_{23}}{C^{00}_{22}}  \right)\right].
\end{split}
\end{equation}

\subsection{pQCD-based Running Gauges}

In previous studies \cite{Xu:2008av, Ferini:2008he, Plumari:2012ep, MacKay:2022uxo, Zhang:1997ej, Wang:2021owa,  Molnar:2001ux,  Lin:2004en}, numerical transport models such as the ZPC and AMPT models utilized a differential cross section that strongly depended on the divergent $\hat{t}$-channel in the gluon-gluon scattering amplitude. To regularize the singularities in divergent channels, the Debye screening mass $m_D$ is introduced \cite{Arnold:2003zc}:
\begin{equation}
\frac{1}{\hat{t}^2}\rightarrow\frac{1}{(\hat{t}-m_D^2)^2}.
\end{equation}

At the extreme temperatures of the QGP, thermal effects modify the screening properties and alter the gauge coupling strength due to temperature-induced momentum transfer. Consequently, the Debye mass $m_D$ and the QCD gauge coupling $g=\sqrt{4\pi\alpha_s}$ become temperature-dependent running parameters. As a running parameter, the QCD gauge coupling $g(T)$ is given as follows \cite{Csernai:2006zz}:
\begin{equation}\label{rung}
\frac{1}{g^2(T)}=\frac{9}{8\pi^2}\ln\left(\frac{T}{\Lambda_T}\right)+\frac{4}{9\pi^2}\ln\left[2\ln\left(\frac{T}{\Lambda_T} \right) \right],
\end{equation}
where $\Lambda_T=30$ MeV is a scale parameter related to the QCD scale $\Lambda_{\mathrm{QCD}}$. The strong coupling constant $\alpha_s$ is then defined as $\alpha_s(T)=g^2(T)/4\pi$. The Debye screening mass is proportional to the product $gT$ in an $N_f$-flavor SU(3) theory \cite{Arnold:2003zc}:
\begin{equation}\label{debye}
m_D^2=\frac{1}{3}\left(3+\frac{1}{2}N_f \right)g^2T^2,
\end{equation}
where $g\sim m_D/T$ is defined as $\left[1/g^2(T) \right]^{-1/2}$ via Eq. (\ref{rung}). These temperature-dependent parameters play a crucial role in determining the interaction strengths and screening effects within the parton medium, and how they affect the shear viscosity of the $N=3$ QGP.

\section{Results} 

In this study, only the divergent scattering channels for all scattering processes are considered in the $N=3$ QGP. These divergent channels are screened with the running Debye mass, and the differential cross sections are calculated using the running QCD coupling: 
\begin{equation}
\begin{split}
&\frac{d\sigma_{11}}{d\hat{t}}\simeq\frac{9\pi\alpha_s^2}{2}\frac{1}{(\hat{t}-m_D^2)^2},~~~~~\frac{d\sigma_{12}}{d\hat{t}}\simeq2\pi\alpha_s^2\frac{1}{(\hat{t}-m_D^2)^2} ,\\\\
&\frac{d\sigma_{22}}{d\hat{t}}\simeq\frac{8\pi\alpha_s^2}{9}\frac{1}{(\hat{t}-m_D^2)^2},~~~~~\frac{d\sigma_{23}}{d\hat{t}}\simeq\frac{8\pi\alpha_s^2}{9}\frac{1}{(\hat{t}-m_D^2)^2}.
\end{split}
\end{equation}
As shown above, the differential cross sections for identical and different fermions are indistinguishable if only the $\hat{t}$-channel exchange is considered. While $d\sigma_{22}/d\hat{t}=d\sigma_{23}/d\hat{t}$ due to identical kinematics, $C^{00}_{22}\neq C^{00}_{23}$ due to differences in Kronecker delta factors that account for species-specific interactions. 
  
For a QGP, the molar fractions under a MB statistical number density are given by
\begin{equation} \label{wk15gqx}
\begin{split}
x_g=\frac{n_g}{n_g+n_q+n_{\bar{q}}},~~~x_q=x_{\bar{q}}=\frac{n_{q,\bar{q}}}{n_g+n_q+n_{\bar{q}}},~~~\mathrm{where}~~~n_j=d_j\frac{T^3}{\pi^2},
\end{split}
\end{equation}
with $d_g=16$ and $d_{q,\bar{q}}=6N_f$ as the gluon and (anti-)quark degeneracy factors, respectively. Thus, $n_g+n_q+n_{\bar{q}}\propto 4(4+3N_f)$ Essentially, the molar fractions $x_g$ and $x_{q,\bar{q}}$ are determined by the ratios of the degeneracies of the individual species to the total degeneracy of the QGP. The flavor number $N_f$ contributes directly to the enhancement of the (anti-)quark population, even though the (anti-)quarks are approximated as a single generic (anti-)quark under the assumption $T\gg m$.

It is important to recognize that quark flavor significally affects the number of species in the QGP by switching the quark population on or off. If no quark flavors are present ($N_f=0$), the QGP simplifies to a gluon gas with $N=1$. However, with nonzero $N_f$, the total number of species includes both gluons and the relevant quark types. For any species number $N$, the number of species in the QGP is 
\begin{equation}
N_{\mathrm{QGP}}= N(1-\delta_{0,N_f})+\delta_{0,N_f},
\end{equation}
where $\delta_{0,N_f}$ is the Kronecker delta distinguishing between the pure gluon gas and $N_f>0$.

\begin{figure}[h!]
\centering
\includegraphics[width=120mm]{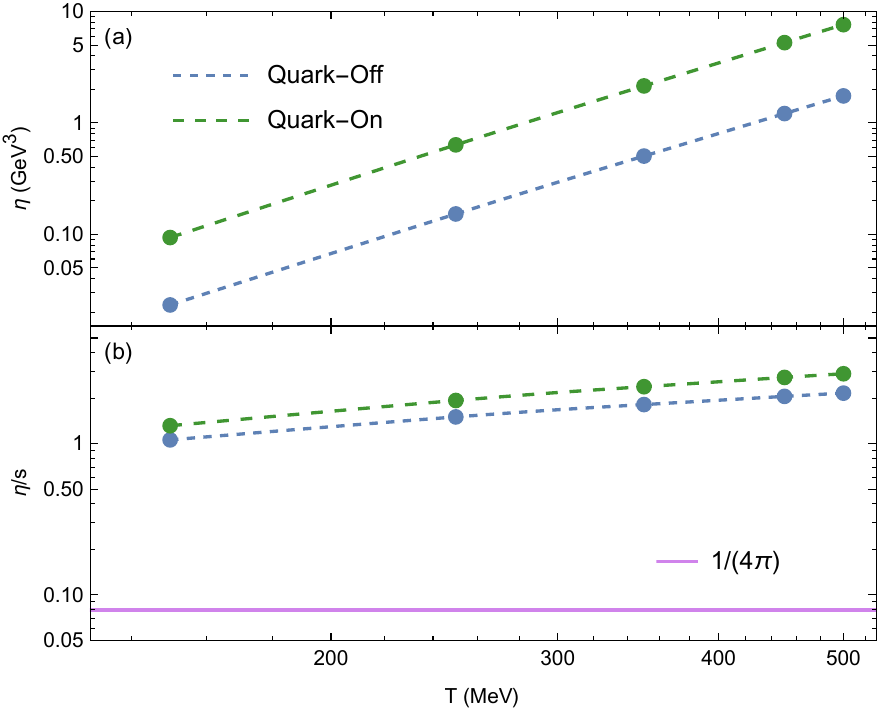}
\caption{\label{eta1} Shear viscosity in GeV$^3$ (a) and unitless $\eta/s$ (b) vs. temperature in MeV. Both plots show the quantites with quarks turned off (blue) and quarks turned on (green). The pink line in panel (b) is the KSS boundary of $\eta/s=1/(4\pi)$.}
\end{figure}

 In this study, we consider three flavors of quarks (up, down, and strange). Figure \ref{eta1}(a) shows the temperature dependence of shear viscosity with the quarks ``switched off'' ($N_f=0$, blue curve) and ``switched on'' ($N_f=3$, green curve). Figure \ref{eta1}(b) plots the corresponding shear viscosity-to-entropy density ratio $\eta/s$ versus temperature, where $s=4(n_g+n_q+n_{\bar{q}})$; the pink line in panel (b) represents the KSS bound of $\eta/s=1/(4\pi)$. 

\section{Discussion}

\begin{figure}[h!]
\centering
\includegraphics[width=120mm]{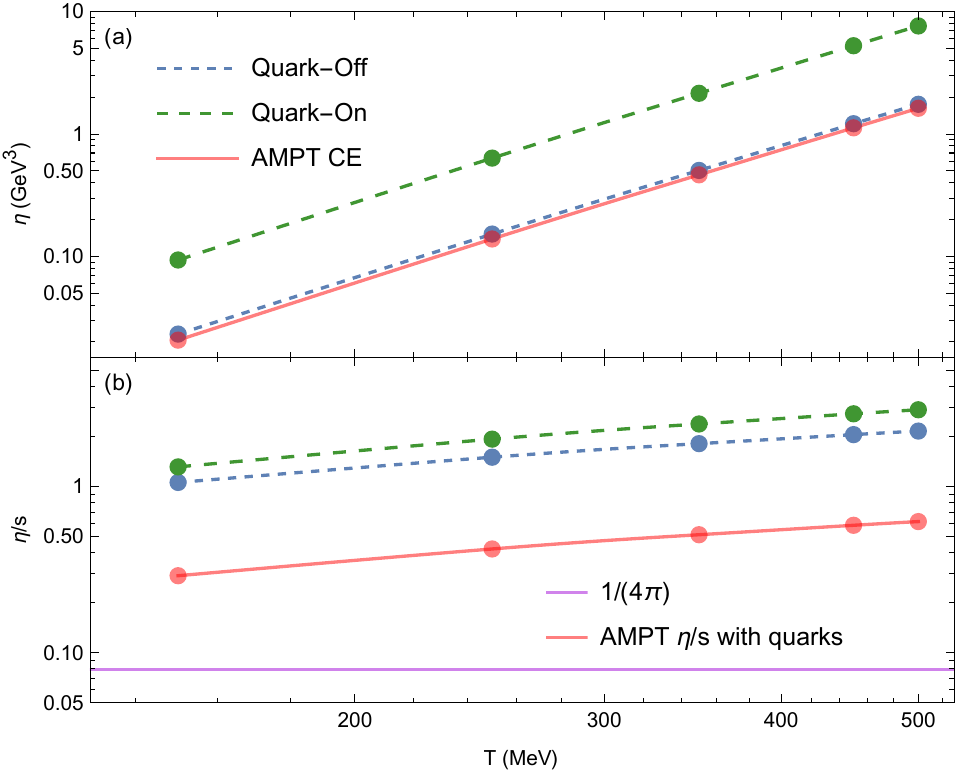}
\caption{\label{eta2} Same as Figure \ref{eta1}, however with one-component formulae for $\eta$ and $\eta/s$ using the AMPT differential cross section in red. The red $\eta/s$ curve corresponds tor $N=1$ AMPT viscosity and $N=3$ entropy density. }
\end{figure}

As observed in Figure \ref{eta1}, the inclusion of quark flavors significantly increases both shear viscosity and entropy density, demonstrating that the shear viscosity is enhanced when the partial viscosities of species-specific interactions are taken into account. This result contrasts sharply with one-component descriptions of QGP shear viscosity and $\eta/s$ in the AMPT approach to QGP, as seen in Figure \ref{eta2}. The AMPT approach to QGP utilizes the following differential cross section to model parton interactions:
\begin{equation}
\begin{split}
\frac{d\sigma_{\mathrm{AMPT}}}{d\hat{t}}=\frac{9\pi\alpha_s^2}{2}\left(1+\frac{m_D^2}{\hat{s}} \right)\frac{1}{(\hat{t}-m_D^2)^2},
\end{split}
\end{equation}
which defines the total cross section as independent of $\hat{s}$: $\sigma_{\mathrm{AMPT}}={9\pi\alpha_s^2}/({2m_D^2})$. 

This formalism oversimplifies QGP scattering dynamics by treating the strong coupling and Debye mass as constant parameters, resulting in a constant $\sigma_{\mathrm{AMPT}}$ \cite{MacKay:2022uxo, mackay1, Lin:2014tya}. Such assumptions neglect the critical temperature dependence of these parameters, which is essential for capturing the non-equilibrium behavior of the QGP. Figure \ref{eta2} illustrates the impact of incorporating a running coupling and a running Debye mass in AMPT, showing how these adjustemtns affect the resulting $\eta$ and $\eta/s$ curves. 

\subsection{Time-dependent cooling}

Assuming the $N=3$ QGP is in full equilibrium (i.e., gluons and (anti-)quarks are perfectly thermalized with the system), $\eta$ and the entropy density $s$ are synchronously influenced by the time-dependent cooling QGP temperature. As described in Eqs. (\ref{rung}) and (\ref{debye}), a decreasing temperature leads to the expansion of the QGP, accompanied by changes in the relevant interaction cross sections due to the running Debye mass and QCD coupling.

\begin{figure}[h!]
\centering
\includegraphics[width=125mm]{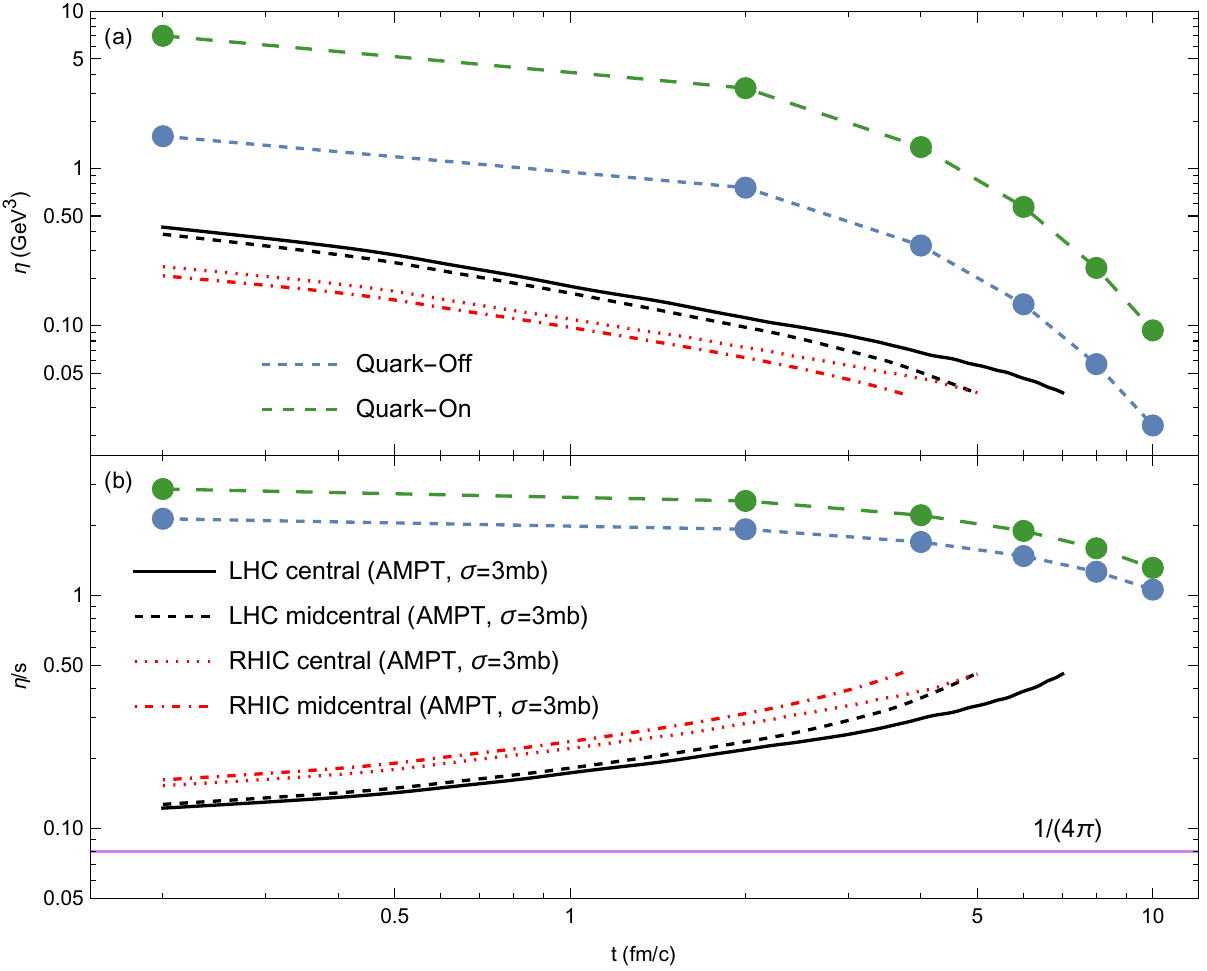}
\caption{\label{eta3} Shear viscosity in GeV$^3$ (a) and unitless $\eta/s$ (b) vs. time in fm/$c$. Like Figures \ref{eta1} and \ref{eta2}, blue curves correspond to quarks turned off and green curves to quarks turned on; the pink line in panel (b) is $\eta/s=1/(4\pi)$. The original AMPT data for (mid-)central collisions from LHC (black) and RHIC (red) are provided for comparison. }
\end{figure}

The cooling of the thermalized QGP can be modeled using Newton's cooling law, assuming an exponential decay of the temperature over time. This approximation is consistent with the rapid expansion and cooling dynamics observed in heavy-ion collisions \cite{J.Adams:STAR, PHENIX:2004vcz, Romatschke:2007mq, Song:2008si, ALICE:2010suc}. Using the boundary conditions $T=550$ MeV at $t=0$ fm/$c$ and $T=150$ MeV at $t=10$ fm/$c$, the temperature as a function of time is given by
\begin{equation}
T(t)=550~\mathrm{MeV}\cdot\exp\left(-\frac{1.204c}{\mathrm{fm}}t \right).
\end{equation}
As the temperature decreases, the QGP expands, with interaction cross sections increasing due to the running parameters $m_D$ and $\alpha_s$. This evolution directly impacts $\eta$ and $\eta/s$ over time; to illustrate these time-dependencies, $T(t)$ from the cooling model is substituted into the CE formulae, producing the trends shown in Figure \ref{eta3}. 

Figure \ref{eta3}(a) reveals that the multi-component viscosity decreases over time as the QGP expands and cools, with the ``quarks on'' case (green curve) consistently exhibiting higher viscosity than the ``quarks off'' case (blue curve). Similarly, Figure \ref{eta3}(b) shows a decrease in multi-component $\eta/s$ over time, approaching the KSS lower bound (pink line). For comparison, the original AMPT measurements of $\eta$ and $\eta/s$ assuming full equilibrium and a constant cross section ($\sigma_{\mathrm{AMPT}}=3\,\mathrm{mb}$) are included for central and mid-central collisions at LHC (Pb+Pb at 2760$A$ GeV, black curves) and RHIC (Au+Au at 200$A$ GeV, red curves). The AMPT measurements used the $N=1$ CE viscosity and a $N=3$ entropy density for (anti-)quarks and gluons \cite{MacKay:2022uxo}. 

As originally discussed, the AMPT results show ``wrong'' trends preferred to Bayesian analysis of experimental data and perturbative QCD, particularly at lower temperatures and later times. This discrepancy arises because AMPT assumes a constant total cross section, independent of energy and temperature, and neglects the contributions of running QCD parameters \cite{MacKay:2022uxo}. Therefore, the multi-component approach aligns more with perturbative QCD, driven by the utilization of running Debye mass and running gauge coupling.

\section{Conclusion}

In this report, the shear viscosity and the shear viscosity-to-entropy density ratio of the quark-gluon plasma were analytically studied using a novel multi-component approach. This method emphasized the critical role of species-specific kinematic contributions to QGP shear viscosity, the incorporation of temperature-dependent running parameters to $\hat{t}$-channel-dominated differential cross sections, and the time-dependence of temperature assuming a fully thermalized system. By comparing these results to established transport models, such as AMPT, we identified significant discrepancies and demonstrated the importance of a multi-component framework for accurately modeling QGP dynamics.

The key findings included a comprehensive temperature-dependent analysis of $\eta$ and $\eta/s$ for an $N=3$ QGP, influenced by species-specific interactions between gluons and (anti-)quarks. These results underscore the limitations of static models, which fail to capture the non-equilibrium properties of the QGP \cite{MacKay:2022uxo}. Moreover, the inclusion of (anti-)quarks in the analysis revealed an enhancement of $\eta$ and $\eta/s$ due to additional partial viscosites from same-species and cross-species interactions. This contrasts with the effective one-component gluon gas treatment used in the ZPC and AMPT models \cite{Xu:2008av, Ferini:2008he, Plumari:2012ep, MacKay:2022uxo, Zhang:1997ej, Wang:2021owa,  Molnar:2001ux,  Lin:2004en}, which oversimplifies the system and underestimates shear viscosity and $\eta/s$. Additionally, using a time-dependent cooling model for the fully thermalized QGP, we demonstrated that both $\eta$ and $\eta/s$ decrease over time as the system expands and cools. This behavior reflects the increasing fluidity of the QGP and aligns with experimental observations of collective flow in heavy-ion collisions.

These findings highlight the necessity of adopting a multi-component framework when studying QGP transport properties, particularly in scenarios involving significant temperature evolution and phase transitions. Our approach provides an optimistic description of the interplay between gluons and (anti-)quarks in the deconfined parton phase, as well as the influence of running QCD parameters on shear viscosity. Further studies could explore the inclusion of AMPT-like corrections to the differential cross sections to account for energy-independent (but running) total cross sections, the effects of chemical non-equilibrium and partial fugacities, and further detailed comparisons to experimental observables such as anisotropic flow coefficients from RHIC and LHC.

\end{document}